\documentclass[11pt]{article}
\usepackage{hyperref}
\pdfoutput 1

\begin{document}
\title{High-Frame-Rate Oil Film Interferometry}

\author{Jonathan C. White, Russell V. Westphal, John Chen\\
\\\vspace{6pt} Mechanical Engineering Department,\\California Polytechnic State University, San Luis Obispo, CA 93407, USA}

\maketitle

\begin{abstract}
The fluid dynamics video to which this abstract relates contains visualization
of the response of a laminar boundary layer to a sudden puff from a small hole.
The boundary layer develops on a flat plate in a wind tunnel; the hole is located
at a streamwise Reynolds number of 100,000.  The visualization of the boundary
layer response is accomplished using interferometry of a transparent, thin
film of oil placed on the surface immediately downstream of the hole and 
with its leading edge perpendicular to the direction of flow.  Through lubrication
theory, it is understood that the rate of change of the spacing of the interference
fringes is proportional to the skin friction at any instant.  For reference, a small
disk-shaped protrusion of the type often used to trip the boundary layer in wind
model tunnel testing is also shown.  Three cases with different puff strengths
are included.  Using a high-speed commercial camera, frame rates in excess of
1000/sec have been recorded; the video shown here was taken at 24 frames/sec
to remain within prescribed file size limits.
\end{abstract}

\end{document}